**Nonwoven Reinforced Photocurable Poly(glycerol sebacate)-Based Hydrogels**

Michael Phillips[1], Giuseppe Tronci[1], Christopher M. Pask[2] and Stephen J. Russell[1]

[1] Clothworkers' Centre for Textile Materials Innovation for Healthcare, Leeds Institute of Textiles & Colour, School of Design, University of Leeds, UK

[2] School of Chemistry, University of Leeds, UK

**Abstract**

Implantable hydrogels should ideally possess mechanical properties matched to the surrounding tissues to enable adequate mechanical function while regeneration occurs. This can be challenging, especially when degradable systems with high water content and hydrolysable chemical bonds are required in anatomical sites under constant mechanical stimulation, e.g. a foot ulcer cavity. In these circumstances, the design of hydrogel composites is a promising strategy to provide controlled structural features and macroscopic properties over time. To explore this strategy, the synthesis of a new photocurable elastomeric polymer, poly(glycerol-*co*-sebacic acid-*co*-lactic acid-*co*-polyethylene glycol) acrylate (PGSLPA), is investigated, along with its processing into UV-cured hydrogels, electrospun nonwovens and fibre-reinforced variants, without the need for a high temperature curing step or use of hazardous solvents. The mechanical properties of bioresorbable PGSLPA hydrogels were studied with and without electrospun nonwoven reinforcement and with varied layered configurations, aiming to determine the effects of microstructure on bulk compressive strength and elasticity. The nonwoven reinforced PGSLPA hydrogels exhibited a 60 % increase in compressive strength and an 80 % increase in elastic moduli compared to fibre-free PGSLPA samples. Mechanical properties of the fibre-reinforced hydrogels could also be modulated by altering the layering arrangement of the nonwoven and hydrogel phase. The nanofibre reinforced PGSLPA hydrogels also exhibited good elastic recovery, as evidenced by hysteresis in compression fatigue stress-strain evaluations showing a return to original dimensions.

Photoresponsive; Nanofibre; Electrospun; UV-Cured; Nonwoven; Regenerative Medicine; Structural Support Scaffold; Tissue Engineering; Biomaterial

## 1. Introduction

Hydrogels are widely applied to support the repair and regeneration of soft tissues, though their large water content can limit clinical handling, fixability to surrounding tissues and wet state mechanical properties. To overcome these issues, non-hydrophilic alternatives may be preferred over systems with enhanced regenerative functionality. One example is in the clinical treatment of ulcerative chronic wounds of the lower feet, specifically cavities caused by tissue loss in the plantar fat pad. Currently, repeated silicone injections can be used to fill such cavities to restore a degree of mechanical function, but the treatment provides no regenerative capacity. In situations such as this, the ability to switch to a bioresorbable elastomeric hydrogel, with appropriate load-bearing properties, could provide scope for both regenerative as well as mechanical function.

In seeking suitable injectable alternatives, poly(glycerol sebacate) (PGS) is an elastomeric, bioresorbable polymer that has been previously deployed in a variety of biomedical applications, and possesses a valuable combination of physical and biological properties. Harding et al.[1] produced PGS based patches for delivery of embryonic stem cells to the heart, demonstrating that cardiomyocytes remain active on the plate for longer than three months until interrupted. PGS has also been investigated for elastomeric scaffolds used for small diameter bypass grafts, and potential advantages have been identified in relation to elastin expression.[2] Scaffolds containing PGS are reported to promote proliferation and phenotypic protein expression in relation to vascular cells.[3] PGS-based materials therefore offer potential to develop biocompatible elastomeric materials, combined with regenerative function for the repair and regeneration of soft tissues.

To further improve the bulk mechanical properties of PGS, particularly when delivered in the form of a hydrogel, micro- or nanofibre reinforcement can be harnessed. Nanofibrous nonwovens are engineered fibrous assemblies containing fibres with diameters up to ~500 nm.[4] Previously, microfibre reinforced hydrogel scaffolds have been produced by melt-electrowriting (MEW) processes, among other methods.[5] Huang et al. demonstrated that integration of nanofibres into an alginate-based hydrogel yielded mechanical behaviour consistent with native tissue, with the compressive stress testing revealing a 'J-curve' response.[6] The stress increased by 87 % with a 30 % nanofibre content. Electrospun gelatin nanofibres have also been used to reinforce alginate hydrogels in scaffolds for corneal tissue engineering, with mechanical properties an order of magnitude higher than for the hydrogel materials alone.[7] Martin and Youssef [8] also studied the dynamic properties of hydrogels with respect to load-bearing for biomedical applications.

Previous work on the production of PGS nanofibres has shown promise for tissue engineering purposes. Gultekinoglu et al.[9] blended a PGS prepolymer with poly(vinyl alcohol) (PVOH), blending the un-crosslinked PGS prepolymer with PVOH avoids solubility and melting issues associated with preparing electrospinning solutions of crosslinked PGS. The resulting fibres had a uniform size distribution, which was maintained even after thermal crosslinking. After washing to remove PVOH, the average fibre diameter decreased by approximately 25%. The study characterised the chemical structure, morphology, and cell viability of the PGS fibres, demonstrating that they supported cell adhesion and proliferation without toxicity over 7 days of interaction. Hou et al.[10] produced microfibrous core-shell fabrics made of PCL and PGS using a wet-wet coaxial electrospinning technique. The researchers immobilised heparin on the surface of these fibres and evaluated their chemical, mechanical, and biological properties. The introduction of PGS in the composite fibres allowed for increased degradability and mechanical properties, combining structural integrity from slowly degrading PCL and fibre elasticity from PGS. The addition of PGS and heparin improved the attachment and proliferation of endothelial cells, making these core-shell fibres promising for tissue-engineering applications. Salehi et al.[11] produced aligned nanofibres composed of PGS and PCL for the purpose of cornea tissue engineering. The fibres were produced via electrospinning using varying weight ratios of PGS and PCL, resulting in diameters ranging from 300 to 550 nm. Analysis showed that increased PGS content decreased overall crystallinity and elastic modulus, while the surface modulus exceeded the elastic modulus and increased with PGS content. This was thought to be due to the increasing PGS content forcing PCL into confined and cross-linked domains near the fibre surface.

Luginina et al.[12] investigated the suitability of PGS/PCL polymers for creating composite fibres incorporating bioactive glass (BG) particles. The researchers produced composite electrospun fibres with BG particles and characterised them. The addition of PGS increased the average fibre diameter, while the presence of BG particles in the composite fibres slightly broadened the diameter distribution without significantly changing the average diameter. The fibres were found to be hydrophilic, and BG particles did not affect fibre wettability. Degradation and acellular bioactivity tests revealed the release of BG particles from the composite fibres, which could be beneficial for therapeutic applications like wound healing. However, the weak interface between BG particles and the polymeric fibres did not lead to improvements in mechanical properties. Preliminary biological tests showed promise for potential use in soft tissue engineering applications.

Despite previous reports on composite fibres made of PGS, fibre-reinforced PGS-based hydrogels are yet to be extensively explored and such design could be leveraged to enhance the mechanical properties of the PGS phase. Other than the mechanical properties, one of the drawbacks of PGS for clinical applicability is that crosslinking the polymer requires high temperature, which prevents any potential clinical treatment in which the polymer is injected directly into a defect site, e.g., a foot ulcer cavity, and cured *in situ* through external stimulus (Figure 1). The ability to cure *in situ* through safe external stimuli could potentially enable less invasive key-hole clinical procedures, as well as enable cavity defects, e.g., foot ulcer cavity, to be more effectively filled with an injectable material to ensure full conformance to the wound surface. This surface interaction could also improve fixation, as well as the structural integrity and durability of the hydrogel following implantation. Photo-curable variants of valuable biocompatible polymers such as collagen have previously been reported [13-16] and have the potential to simplify clinical procedures.

Therefore, to extend the range of clinical applications for PGS and remove some of the current limitations, we report the synthesis of a photocurable PGS co-polymer, i.e. poly(glycerol-*co*-sebacic acid-*co*-lactic acid-*co*-polyethylene glycol) acrylate (PGSLPA), and its processing into UV-cured hydrogels and electrospun fibres.

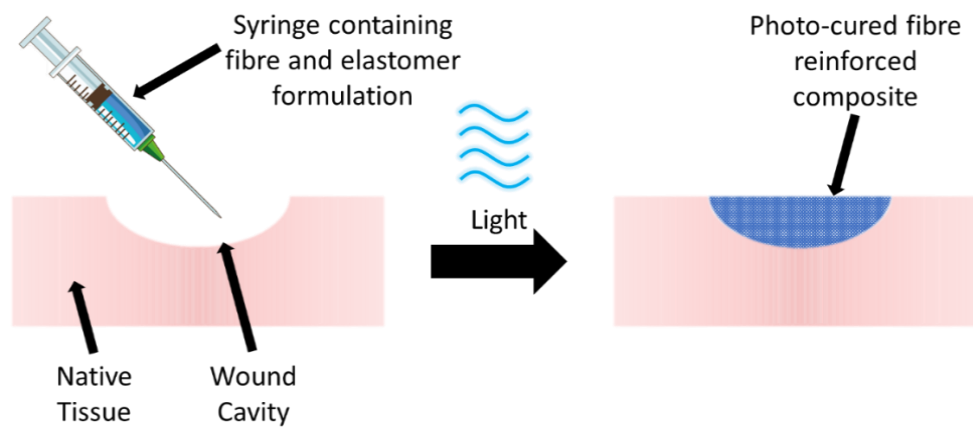

**Figure 1.** Potential clinical treatment in which a PGS-based polymer formulation containing reinforcing fibres is injected into a defect such as a wound cavity to promote repair and regeneration. Following *in situ* irradiation (photocuring) of the formulation at an appropriate wavelength, (e.g., blue light), the fibre reinforced PGS-based hydrogel is formed.

We also study the properties of PGSLPA hydrogels reinforced with electrospun nanofibres made of PGSLPA and poly($\varepsilon$-caprolactone) (PCL), to determine prospects for modulating mechanical properties, including elastomeric response. The building blocks of the hydrogel composite were selected aiming to de-risk the material's potential use in the biological environment. Glycerol, lactic acid, PCL and Polyethylene glycol (PEG) exhibit low cytotoxicity and hold FDA approval for various applications such as dermal fillers.[17] Extensive washing and dialysis were also carried out on the final products to ensure purification and removal of potentially toxic compounds used during either the synthesis or composite manufacture.

**2. Materials and Methods**

*2.1. Materials*

Glycerol, 4-dimethylaminopyridine, triethylamine, were obtained from Fisher Scientific. Sebacic acid was obtained from Alfa Aesar. Dichloromethane, polyethylene glycol ($M_w$ 8000), stannous octanoate, p-toluenesulfonic acid monohydrate, lactic acid (85 %), polycaprolactone ($M_w$ 80000), 1,1,1,3,3,3-hexafluoro-2-propanol (HFIP), polyethylene glycol diacrylate ($M_w$ 8000) (PEGDA$_{8000}$), 2-hydroxy-4′-(2-hydroxyethoxy)-2-methylpropiophenone (Irgacure 2959) and anhydrous magnesium sulphate (MgSO$_4$) were obtained from Sigma Aldrich.

*2.2. Synthesis of Poly(glycerol sebacate-co-lactic acid-co-polyethylene glycol) (PGSLP)*

Poly(glycerol sebacate-*co*-lactic acid-*co*-polyethylene glycol) (PGSLP) was synthesised as a starting material by adapting the method reported by Jia et al.[18] First, a round bottom flask was charged with sebacic acid (11.0 g, 54.35 mmol), lactic acid (14.67 g, 163.04 mmol) and PEG (M$_w$ 8000, 10.87 g, 1.36 mmol). This mass of lactic acid refers to the mass of 85 % solution, taking into account the presence of water. The vessel was flushed with nitrogen, magnetically stirred at 350 rpm and heated to 120°C. Once the temperature reached 120°C, vacuum pressure (0.001 bar) was applied and stirring was continued for 4 h. Stannous octanoate, Sn(Oct)2 (1.1 g, 2.72 mmol) and p-toluenesulfonic acid monohydrate, TSA·H2O (0.52 g, 2.72 mmol) were added, under vacuum (0.001 bar), and the temperature was increased to 180°C. The reaction was then continued for 24 h. Glycerol (5.0 g, 54.35 mmol) was then added, and the reaction continued for a further 8 h under reduced vacuum pressure (5 x 10$^{-4}$ bar). The material produced was a clear viscous fluid. The material was dissolved in DCM, water removed with magnesium sulphate (5 g) and the solvent removed via rotary evaporation. A viscous brown fluid was obtained, that was left for 24 h in a vacuum oven at 40 °C.

*2.3. Synthesis of Photo-functional Poly(glycerol sebacate-co-lactic acid-co-polyethylene glycol) Acrylate (PGSLPA)*

PGSLP (35.0 g) and DMAP (1.66 g, 13.59 mmol) was added to a flame dried round bottom flask, the flask sealed and flushed with N$_2$. Anhydrous dichloromethane (150 mL) was added to the flask and the magnetic stirrer turned on at 350 rpm, the flask was then cooled in an ice bath to 0 °C. Acryloyl chloride (24.46g, 271.74 mmol) and triethylamine (27.45 g, 271.74 mmol) were added dropwise simultaneously using dropping funnels and the reaction left stirring for 24 h. The reaction mixture was poured in ethyl acetate to facilitate precipitation of the polymer, however no precipitate formation occurred. The mixture was placed on a rotary evaporator to remove all the solvent, whereupon the obtained polymer was dissolved in dichloromethane and poured into n-pentane, yielding polymer at the bottom of the flask. The solvent mixture was decanted, the polymer dissolved in dichloromethane and placed on a rotary evaporator. The mixture was dissolved in deionised $H_2O$ and dialysed over a period of 3 days with regular water changes. The polymer solution was then dried on a rotary evaporator yielding a viscous brown fluid.

## 2.4. Preparation of Nonwovens from PGSLP and PGSLPA

Nanofibres of PGSLP and PGSLPA products were electrospun in the presence of PCL as a fibre-forming carrier polymer, to accomplish a suitable degree of chain entanglement for homogeneous fibre formation. A solution of 5 wt % PGSLP (or PGSLPA) and 5 wt % PCL ($M_w$ 80000) was prepared in HFIP, transferred to a glass syringe equipped with a 18G needle tip and loaded onto a syringe pump. A charged plate with a 20-cm distance from the needle tip was secured in place, whereby an electrostatic voltage and syringe flow rate were varied in the range of 20-25 kV and 0.25-1.5 mL $h^{-1}$, respectively, to accomplish homogeneous electrospun fibres. Samples were also produced from electrospinning solutions loaded with photoinitiator (0.2 wt % DMPA).

## 2.5. Scanning Electron Microscopy

SEM micrographs were collected on a Jeol JSM-6610LV scanning electron microscope employing a 10 mm focusing lens. Micrographs were collected at accelerating voltages from 15 to 30 kV and magnification from 250 to 10000. Different accelerating voltages were used to enhance resolution of the nanofibers present in the nonwoven web. Magnification was changed to allow accurate fibre diameter and porosity measurements. Samples were prepared by cutting to 10 x 10 mm, sputter coating was employed to produce high quality micrographs. SEM micrographs were analysed using ImageJ, average fibre diameter was determined by measuring fibre diameters of 50 individual fibres selected at random using the reference scale bar in the SEM micrograph.

## 2.6. Preparation of Photo-Curable Fibre Reinforced PGSLPA Hydrogels

In the fibre-free configuration, a solution of 26 wt % PGSLPA, 10 wt % PEGDA ($M_w$ 8000), and

1 wt % Irgacure 2959 was prepared in deionised water in dark at 50 °C under magnetic stirring at 200 rpm. A Chromato-Vue C-71 ($\lambda$: 365 nm) was used to cure the resulting polymer solution and obtain the gels. The hydrogel-forming solution and respective PGSLPA/PCL nanofibrous webs were layered into a mould in three different composite configurations, as reported in Figure 2.

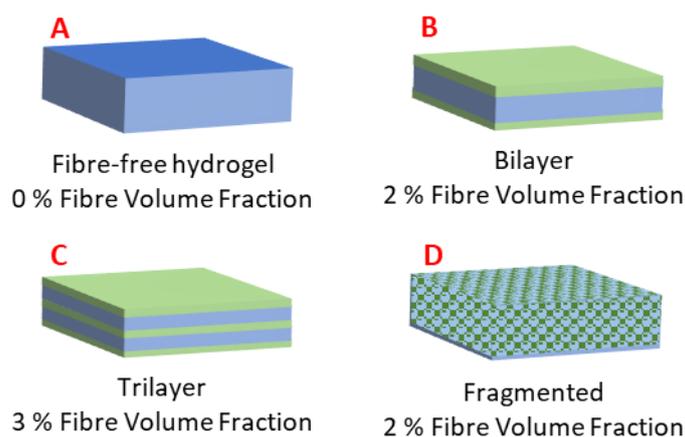

**Figure 2.** Hydrogel and electrospun nonwoven fibre reinforced structural configurations: A) Fibre-free hydrogel of PGSLPA. B) Bilayer arrangement, with the nonwoven layer coated on both sides by the hydrogel. C) Trilayer arrangement via sequentially layered hydrogel and nonwoven. D) Hydrogel reinforced by fragmented nonwoven.

For the fragmented nanofibre-reinforced composite, the electrospun nanofibrous web was cut into 2x1 mm samples and dispersed throughout the hydrogel-forming solution. Moulds were UV irradiated for 2 h to promote photocuring. Figure 2 shows the schematic structural arrangements of the hydrogel and nanofibre components, with the associated volume fractions of the hydrogel (matrix) and nonwoven (reinforcement) phases.

*2.7. Mechanical testing*

Mechanical testing was carried out for the composite constituents, i.e. hydrogel and nanofibre fabric, as well for all composite configurations, in either tension or compression mode, as reported below.

2.7.1. Tensile Testing of Nonwoven Fabrics

Uniaxial tensile testing of nonwovens was conducted (James Heal, Titan Universal Strength Tester), operating with a 100N load cell. Jaw separation was calibrated to 30 mm and checked manually using Preciva FRDM730002 Verniers, with testing performed at an extension rate of

1 mm/min. Specimens were attached to cardboard templates using double sided adhesive tape as anchor points, the effective testing length of the samples was 30 x 25 mm. Once clamped in the jaws, the collection plate was cut from the sample. Sample measurements were done in triplicate, and samples were conditioned at 20 °C and 65 % relative humidity for 24 h prior to testing.

2.7.2. Compression Testing of Hydrogel and Nonwoven Reinforced Samples

Compression tests were conducted to sample failure and the stress-strain response recorded. Photocured samples of PGSLP-based fibres and hydrogels were cut to 4 x 15 x 15 mm prior to testing. An Instron 3365 Universal Tester with a 500 N load cell in compression mode was used, at a strain rate of 1 mm/min and a distance of 2.5 mm. The elastic modulus of the nonwoven reinforced samples were calculated between strain rates 0.1 – 0.6.

*2.8. Degradation Tests*

To simulate physiological conditions, electrospun samples (n=3) were individually incubated in phosphate buffered solution (PBS, 10 mM, pH 7.4, 25 °C) [19] for up to 8 weeks, prior to determination of any mass loss. DMPA photoinitiator doped samples were also tested before and after curing to determine the change in the degradation behaviour due to the introduction of covalent crosslinks at the molecular scale. Samples were cut to equal dimensions (10 x 30 mm), the initial mass of the dry samples was measured, and then re-measured at 1-week intervals, to quantify the percent residual mass. All mass measurements were obtained using oven dried samples (dried at 50 °C for 24 h, in a Binder ED56 Series static oven). Results were linearly fitted to assess erosion-driven degradability. Fitting of a linear model using OriginPro was used to determine the fitting equations and the coefficient of determination ($R^2$) for each degradation profile, allowing an understanding of the relationship between the variables time and mass.

**3. Results and Discussion**

*3.1. Synthesis of PGSLP and PGSLPA*

Synthesis of PGSLP followed a similar methodology to that described by Jia et al.[18]. The polymer was found to be soluble in deionised water with stirring. ¹H NMR analysis (Figure 3) was performed in CDCl₃ using a Bruker Avance III HD 400 MHz spectrometer and confirmed the correct target structure was obtained following synthesis. Peaks at 5.1 – 4.9 ppm correspond to methylene protons from sebacic acid and lactic acid. Peaks at 2.25 ppm were assigned to methylene proton from sebacic acid closest to the carboxylic acid terminal groups. The large peak at 3.55 ppm corresponds to the protons present in the PEG block. Peaks at 4 – 4.4 ppm correspond to methylene protons in glycerol. Peaks at 4.9 – 5.1 ppm were observed due to protons in glycerol and lactic acid. The polymer was further purified by dialysis, leading to a yield of 90%.

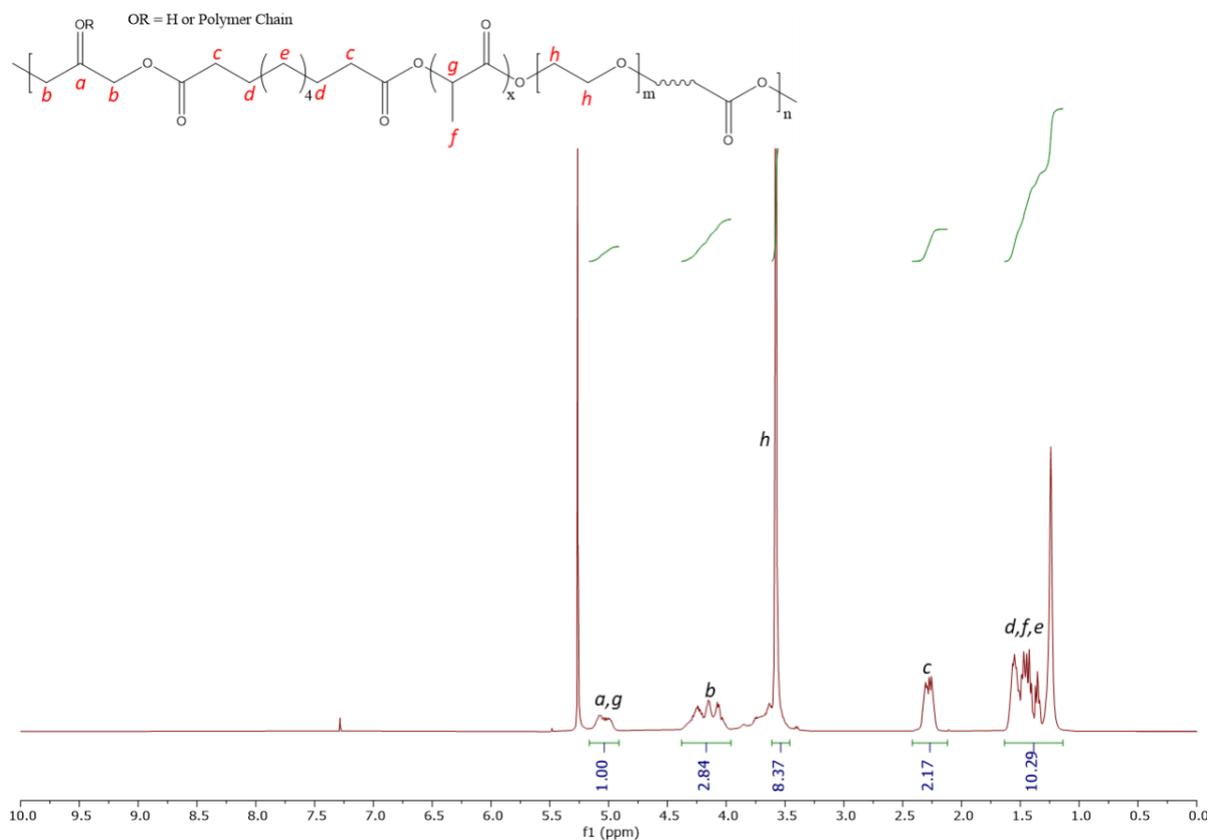

**Figure 3.** ¹H NMR, 300 MHz spectrum (CDCl₃) of PGSLP.

Normally, PGSLP is cured by high temperature treatment, which limits its clinical applicability in cases where curing may need to be done *in situ*. Development of a photocurable derivative was therefore targeted to enable crosslinked network formation at room temperature.

Acryloyl chloride was used to graft alkene functionality, specifically acrylate functional groups to the free hydroxyl groups present within the polymer's structure. Introducing a diacrylate crosslinker, such as PEG diacrylate with a suitable photoinitiator allows for covalent bonds to be formed between the polymer chains. This bond formation allows for a gel structure to form in suitable solvents, which can then be reinforced or doped with suitable fillers.

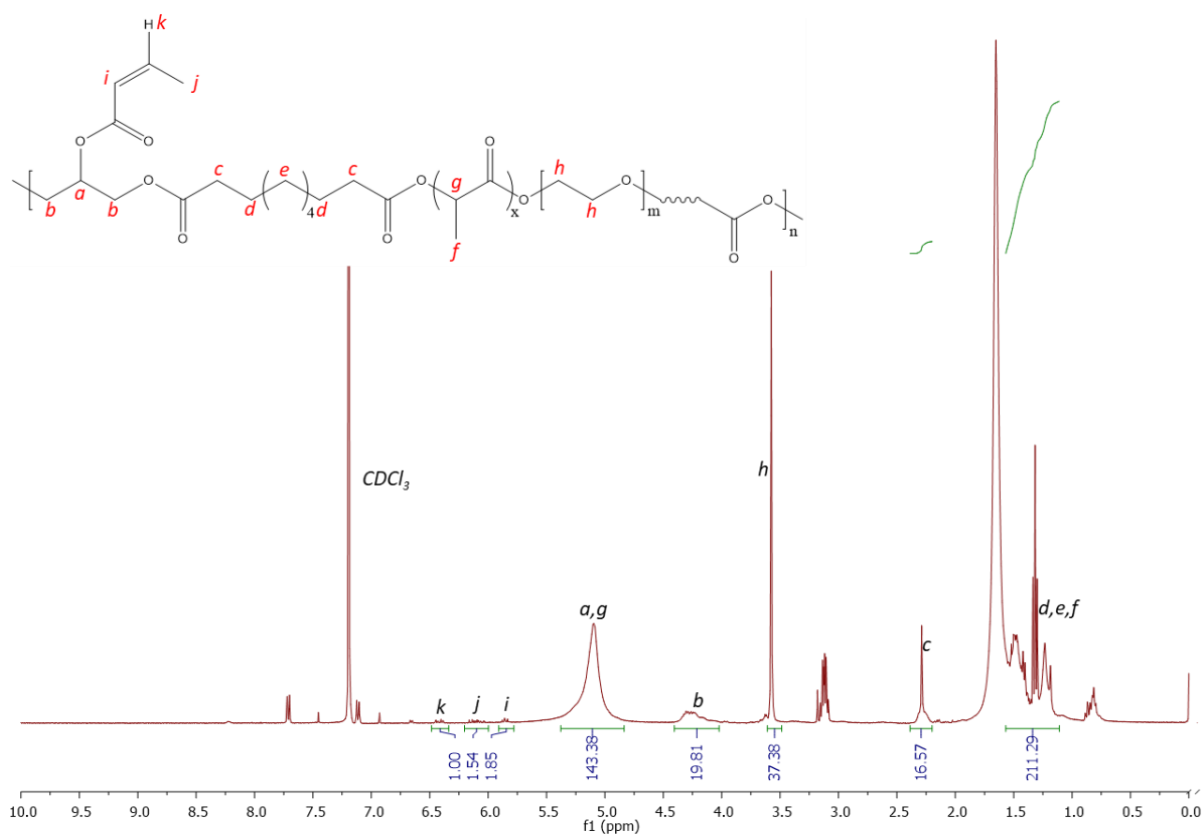

**Figure 4.** $^1$H NMR spectrum 300 MHz (CDCL$_3$) of PGSLPA.

Figure 4 shows the $^1$H NMR spectrum, confirming the successful grafting of acrylate functionality to the polymer structure. A similar peak structure is observed with the unmodified polymer, however additional peaks are observed at 6.4, 6.1 and 5.6 ppm, which correspond to the successfully grafted photocuring functionality. The protons present in the alkene group (labelled *i, j,* and *k*) are clearly observed within the spectrum, corresponding to comparable integration values, as expected. Solubility testing of the material by immersing it in deionised water showed the material to be readily dissolved with stirring, such that it could be an appealing candidate for gel formation *in situ* following injection directly into a wound site and photocuring.

## 3.2. Dimensional, Tensile and Degradation Properties of Electrospun PGSLP and PGLSPA Nonwovens

PGSLP and PGSLPA were successfully electrospun using PCL as a carrier material in 50:50 w/w proportions. This mixing ratio ensured that no electrospraying occurred and the electrospinning process was stable. While commixing of these materials was expected to negatively influence resultant elastic properties of the PGSLP and PGSLPA, polymer blending was identified as a prerequisite for successful electrospinning of self-supporting nonwovens that could subsequently be incorporated into reinforced hydrogel constructs. Electrospun samples were also produced containing photoinitiator (DMPA) to permit subsequent photocuring in the composite structure, avoiding post-spinning incubation in a photoinitiator-supplemented solution.[20, 21] Figure 5 shows typical examples of SEM micrographs of the produced electrospun webs, showing relatively smooth PGSLP/PCL and PGSLPA/PCL fibre morphologies. Table 1 reports the mean fibre diameters of the produced electrospun nonwovens, which were all in the range 280 to 290 nm with comparable standard deviation values.

**Table 1.** Mean fibre diameters of the electrospun nonwovens.

| Sample | Mean Diameter (nm) |
|---|---|
| PCL 100% | 280 ± 94 |
| PGSLP 50% / PCL 50% | 289 ± 94 |
| PGSLPA 50% / PCL 50% | 282 ± 91 |
| PGSLPA 50% /PCL 50%/DMPA 0.2% | 283 ± 91 |
| PGSA 50% /PCL 50%/DMPA 0.2% | 288 ± 92 |

Sufficient polymer chain entanglement, polymer concentration, and solvent distribution over the entangled polymer molecules, enabled the production of smooth, uniform nanofibers. Qian et al.[22] fabricated electrospun PCL and PCL/Chitosan-Gelatin nanofibrous mats and showed that the morphology of the PCL nanofibers was smooth with interconnections between the fibres. This morphology was lost when blending with chitosan-gelatin, with the fibres losing this interconnection, likely being due to an increase in conductivity from the blended solutions from the polar groups associated with chitosan and gelatin, specifically the amine and carboxylic acid functionality. Luginina et al.[12] observed an increase in average fibre diameter when mixing PCL with PGS, likely due to the total increase in volume of polymer content in solution. The average fibre diameter values reported in this work are lower than those reported in the literature for PCL/PGS blends, where the fibre diameters vary in the range of 550 to 4700 nm.[11, 23-25] This could be due to differences in the molecular weights of the PGSLP and PGSLPA produced in this work, compared to previously reported PCL/polymer blends, and the associated impact on the spinning solution viscosity, which will affect as-produced fibre diameters. Differences in the concentration of the spinning solution will also influence the fibre diameter. The solvent choice and the associated volatility, polarity and surface tension are further factors affecting the electrospinning process. Ultimately, the precise electrospinning process parameters and environment in which the fibres are produced will markedly influence resultant mean fibre diameter.

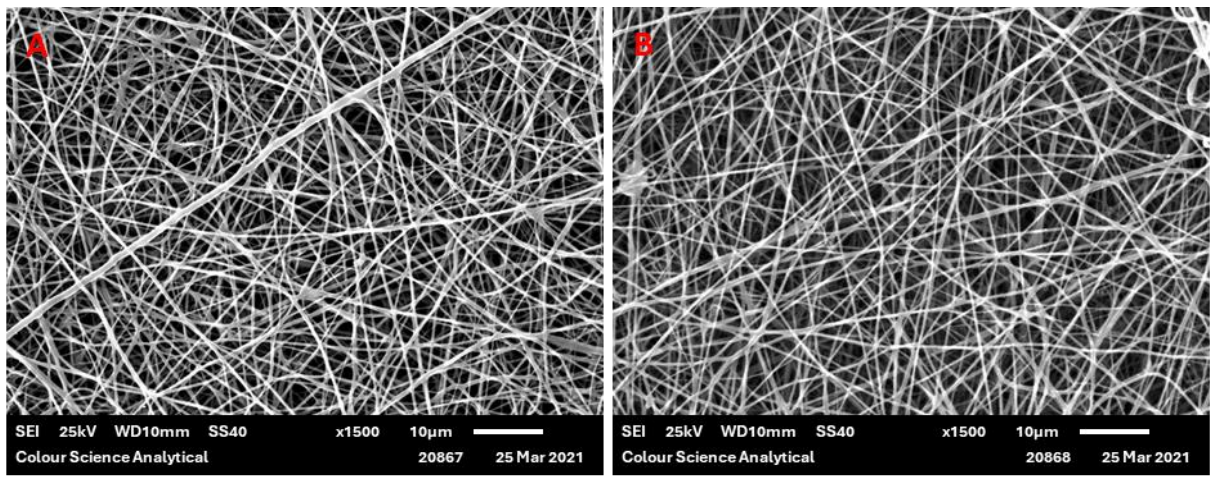

**Figure 5.** SEM micrograph of (A) electrospun 50:50 w/w PGSLP/PCL fibres and (B) electrospun 50:50 w/w PGSLPA/PCL fibres (scale bar = 10 microns).

Stress-strain curves of electrospun materials were collected on a James Heal, Titan Universal Strength Tester. Stress-strain curves of composite and gel materials were collected on a Instron 3365 Universal Tester.

Corresponding stress-strain curves for all PGSLP and PGSLPA electrospun nonwovens are shown in Figure 6, where the measurements of a PCL-based electrospun fabric produced under identical spinning conditions is included as control. PGSLP and PGSLPA-containing nonwovens produced increased toughness in their stress-strain responses compared to the

100% PCL sample (Figure 6). This observation can be explained in terms of composite mechanics and the improved stress distribution due to the elastomer distributing strain more efficiently, reducing localised stress concentrations that may occur in a purely rigid material. The stiffness of the PCL component, when combined with the compliance (flexibility) of the PGSLP and PGSLPA elastomers, leads to an overall reduction in the effective stiffness of the composite. Modification of the molecular structure to introduce photocuring functionality resulted in no significant change in strain response (P<0.05). Electrospun PGSLPA fabrics doped with 0.2 % DMPA photoinitiator displayed no significant change (P<0.05) as compared to undoped electrospun fabric, which may be attributed to the lack of covalent crosslinks and the minimal effect of secondary interactions between grafted residues. On the other hand, UV irradiation revealed a marked change in the stress-strain response, with the fabric displaying a linear stress-strain curve, as a result of covalent crosslinks introduced between PGSLPA polymer chains within the electrospun nanofibres (Figure 6).

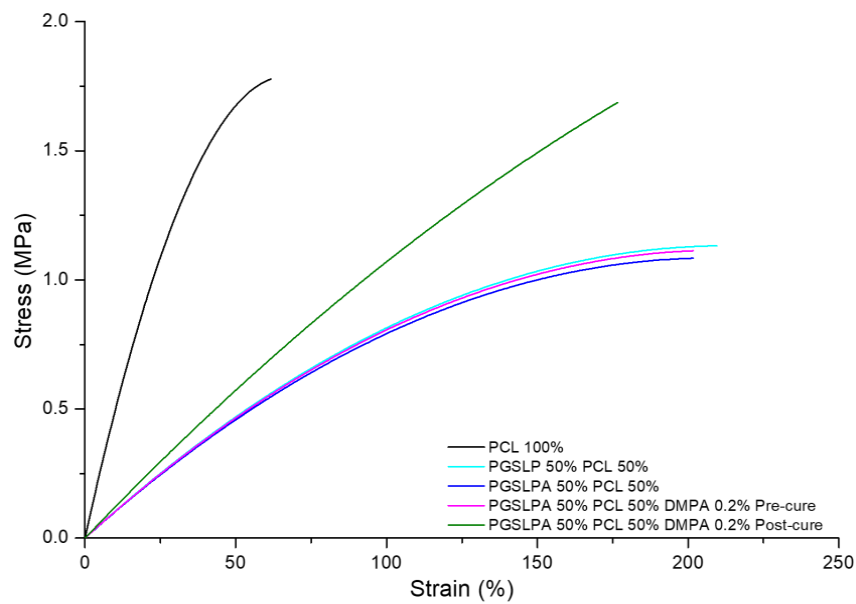

**Figure 6.** Dry state tensile stress-strain curves of PCL control, as well as DMPA-free and DMPA-doped PGSLPA nonwovens before and after UV irradiation.

To reflect physiological conditions, wet tensile properties were also measured (Figure 7), and a marked difference between wet and dry conditions was immediately apparent for the PGSLP and PGSLPA samples.

Other than PCL, the breaking stress and strain of all the hydrated electrospun nonwovens

(Figure 7) were lower than in their dry state (Figure 6), which is a consequence of the water-induced swelling due to the hydrophilicity of the PGSLPA phase.

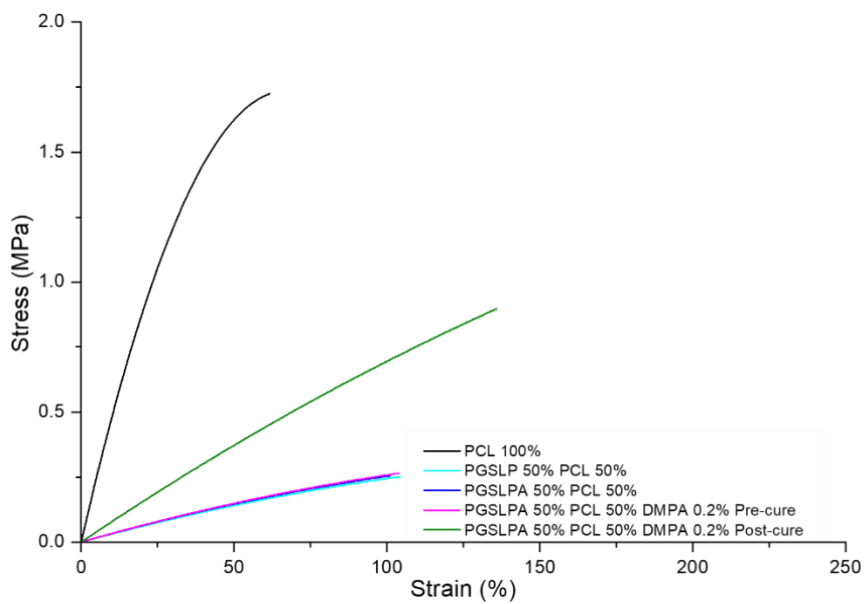

**Figure 7.** Wet state tensile stress-strain curves of PCL control, as well as DMPA-free and DMPA-doped PGSLPA nonwovens before and after UV irradiation.

Photo-curing of PGSLPA reduced the overall decrease in breaking stress (47%) and breaking strain (23%) as compared to the larger decreases prior to photo-curing. This supports the synthesis of UV-induced covalent crosslinks between fibre-forming polymer chains, leading to fibres with increased tensile strength and increased resistance to mechanical deformation.

Cyclical strain experiments (samples loaded to 50 % strain) were carried out to assess the elastic response and revealed no marked hysteresis in the PGSLP and PGSLPA electrospun nonwovens (Figures 8B and 8C) most likely because of the semi-crystalline nature of the PCL present in the fibres [26]. After initial loading (cycle 1), the equilibrated stress-strain response for subsequent cycles closely followed that of the relaxation phase of cycle 1. Mixing of semi-crystalline and amorphous polymers can result in stress concentrations at the interface between the two phases, resulting in poor hysteresis. Herein, the energy dissipated for cycle 1 is 30.4, 5.5, 5.7 and 6.6 for the materials in Figure 8 A, B, C and D respectively. Optimisation of the blend composition and introduction of compatibilisers can enhance the interfacial adhesion and improve the hysteresis response.

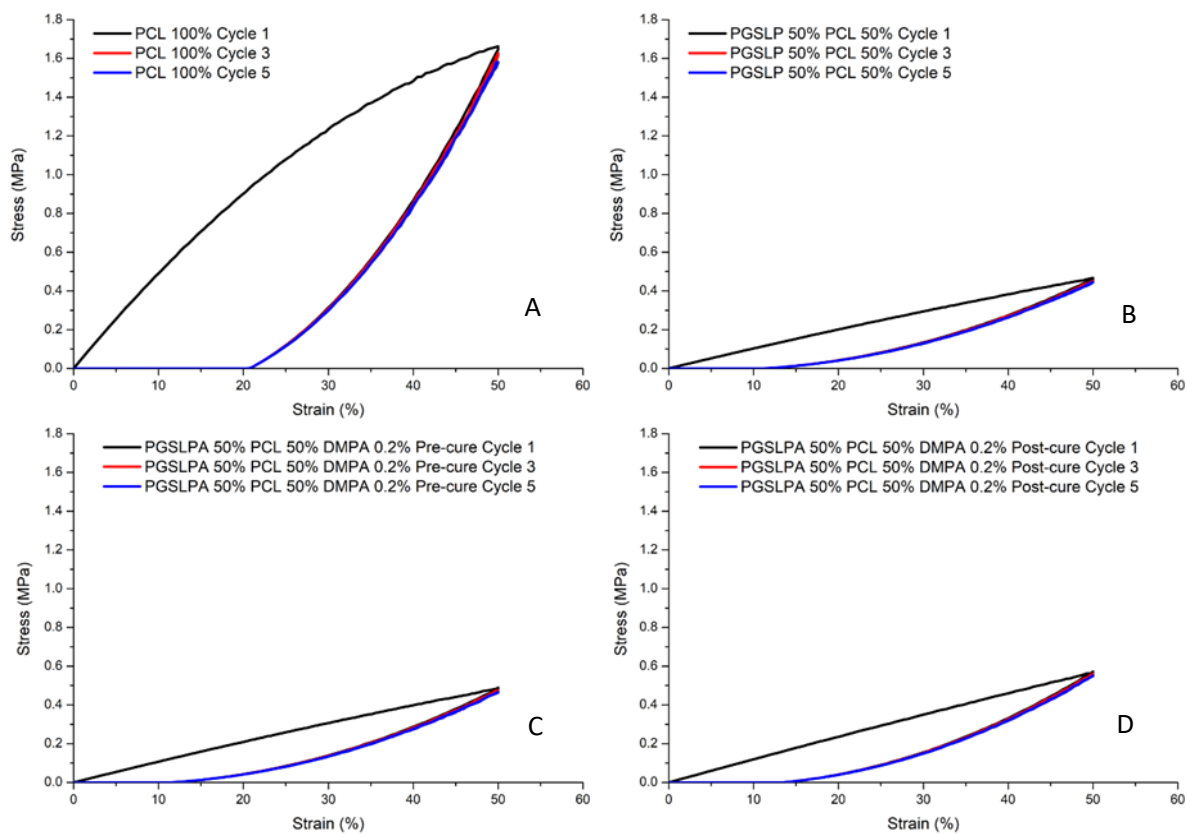

**Figure 8.** Cyclic tensile stress-strain curves (cycles 1, 3 and 5) for: (A) 100 % PCL nanofibre fabrics; (B) 50 % PGSLP/PCL nanofibre fabrics; (C) uncured 50 % PGSLPA/PCL DMPA photoinitiator-doped nanofibre fabrics and (D) cured 50 % PGSLPA/PCL DMPA photoinitiator-doped nanofibre fabrics.

Figures 8C and 8D again reveal increased breaking stress (post photocuring), whereby the same trend of equilibration after the initial loading cycle is observed, with subsequent loading and relaxation phases following the relaxation phase of cycle 1.

Following characterisation of the tensile behaviour, the biodegradability of the electrospun samples was subsequently investigated. Figure 9 shows the mass loss profiles of the nanofibre fabric samples after immersion in PBS, whereby all uncured PGSLP and PGSLPA samples proved to degrade at a higher rate than the 100 % PCL reference sample. Before curing, the presence of photoinitiator did not substantially affect the mass loss, as expected given the absence of covalent crosslinks between fibre-forming polymer chains. However, after photocuring of the photoinitiator-doped PGSLPA fabrics, the sample mass of PGSLPA was retained at a sharply increased level comparable to that of 100 % PCL, with only about 5 % mass loss after 8 weeks. In clinical practice, non-toxic compounds resulting from the breakdown of PGSLP and PGSLPA are easily eliminated from the body. With respect to the application as chronic wound pad materials, this polymer contrasts with currently used silicone injections, where the mechanically degraded material can migrate through the body.[27] On the other hand, PGSLP is comprised of chemicals used in food applications and in some cases is already present in the body (lactic acid), which is intended to minimise potential toxicity issues of the degradation products. However, the post-synthesis modification to yield PGSLPA which has an acrylate functional group, would require detailed toxicology studies to be carried out in future work.

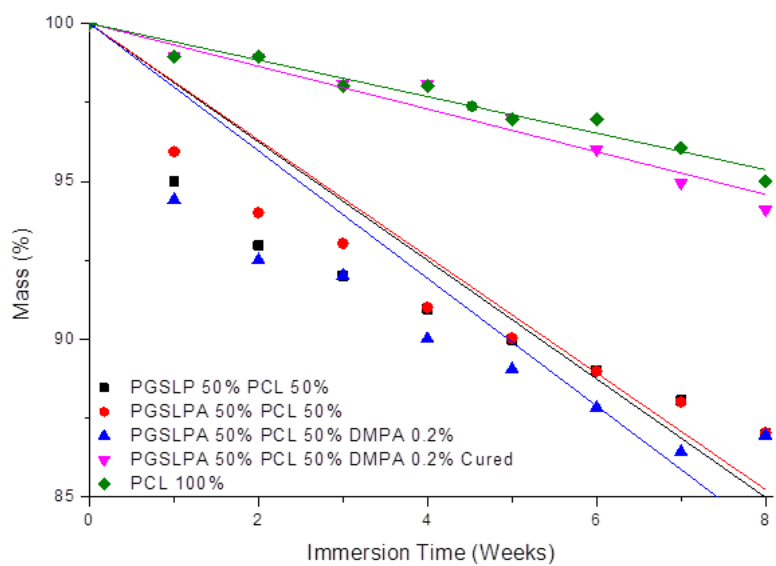

**Figure 9.** Degradation profile of PGSLP and photoinitiator doped PGSLPA based nonwovens before and after photocuring following long-term immersion in PBS buffer solution. The 100 % PCL electrospun nonwoven sample is included as a reference.

The degradation profiles (Figure 9) suggest that the nanofibre fabric samples undergo surface erosion, as there is a controlled mass loss with time (Table 2). The mass loss rate decreases over time indicating that surface erosion causes a reduction in fibre diameter, and as the fibre diameter decreases, the surface area of the fibre therefore reduces, such that less surface erosion can take place. A controlled mass loss is desirable in biomedical implants and injectable formulations, as it allows for retention of mechanical properties as the material is removed by the body and replaced with natural tissue. Another added benefit of a controlled degradation profile is the possibility to couple degradation with release of a therapeutic agent. Incorporation of therapeutic molecules within the structure could provide an avenue for improved clinical outcomes.

Table 2. Linear fitting equations related to degradation profile of nonwovens.

| Nonwoven | Fitting Equation | $R^2$ |
|---|---|---|
| PCL | -0.58x + 100 | 0.9999 |
| PGSLP 50% PCL 50% | -1.88x + 100 | 0.9995 |
| PGSLPA 50% PCL 50% | -1.84x + 100 | 0.9997 |
| PGSLPA 50% PCL 50% DMPA 0.2% | -2.02x + 100 | 0.9994 |
| PGSLPA 50% PCL 50% DMPA 0.2% Cured | -0.68x + 100 | 0.9999 |

PCL is one such material that has previously been investigated for its use as a long-term drug delivery vehicle [28]. Loading of PCL-based fibres with self-assembling peptides has been reported to generate bioactive electrospun meshes with a spider-net nanofiber architecture, due to the renaturation of hydrogen bonds following solvent removal.[29] Likewise, PCL-based, photosensitiser-encapsulated electrospun meshes revealed significantly reduced fibre diameter and significantly increased tensile modulus [30, 31], consequent to the impact of loading on both electrospinning solution viscosity and polymer crystallinity. Further studies should therefore be carried out to investigate the effect of drug loading on either the mechanical properties or the microstructure of the presented hydrogel composite.

*3.3. Compressive Strength of Nonwoven Reinforced PGSLPA Hydrogels*

Compression to failure testing enabled a complete picture of the strain behaviour, indicating how the bulk material would likely perform in a soft tissue repair application. In reality, biomaterials replacing soft tissues are not always required to undergo compression to failure *in situ*, but rather to undergo cyclic compression in the linear, elastic portion of the stress-strain curve. Figure 10 illustrates the ductile stress-strain behaviour of the hydrogel-nonwoven reinforced structural formats (Figure 2 B, C and D) and the non-reinforced hydrogel (Figure 2 A). In all experiments, the nonwoven used to reinforce the hydrogel was the PGSLPA/PCL electrospun sample. Despite the very low fibre volume fraction of only ~2-3% (Figure 2), nanofibre reinforcement increased the maximum stress and modulus of both the PGSLP and PGSLPA hydrogels compared to the non-reinforced samples (Table 3).

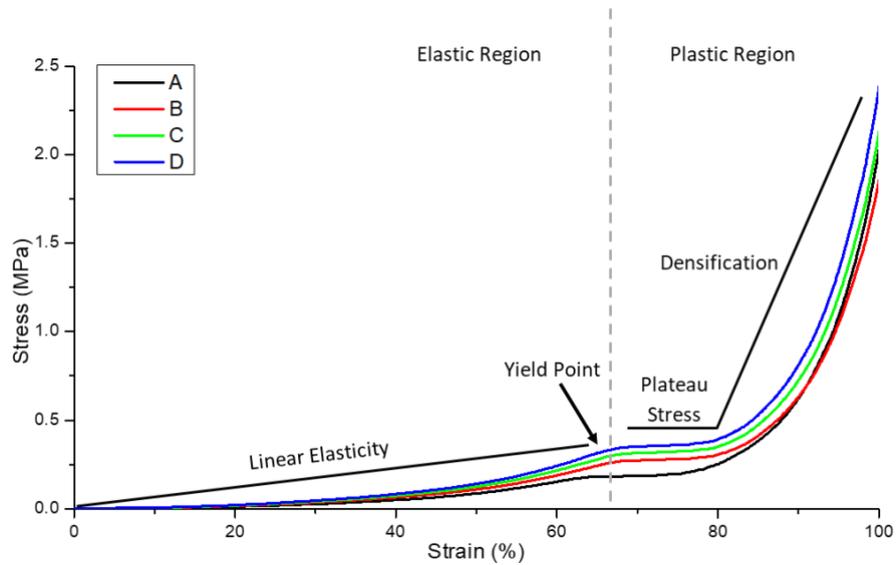

**Figure 10.** Compression stress-strain curve for the hydrogel and nanofibre reinforced PGSLPA hydrogels. A) 100 % PGSLPA hydrogel with no reinforcement. B) Bilayer nonwoven reinforced format. C) Trilayer nonwoven reinforced format D) Segmented nonwoven reinforced format.

The structural format of the hydrogel-nanofibre composite also appeared to influence the compression stress-strain response. In Figure 10 the highest stress was associated with the segmented nonwoven reinforced sample (Figure 2 D). For example, at 60 % strain, the stress of this sample was 59 %, 29 % and 12% greater than that of an unreinforced hydrogel (Figure 2 A), bilayer (Figure 2 B) and trilayer sample (Figure 2 C) respectively.

Figure 11 shows the linear portion of the stress-strain curve profile for all four compositions, i.e., before the yield point, and this was considered for fatigue testing, as well as to calculate the elastic modulus (Table 3).

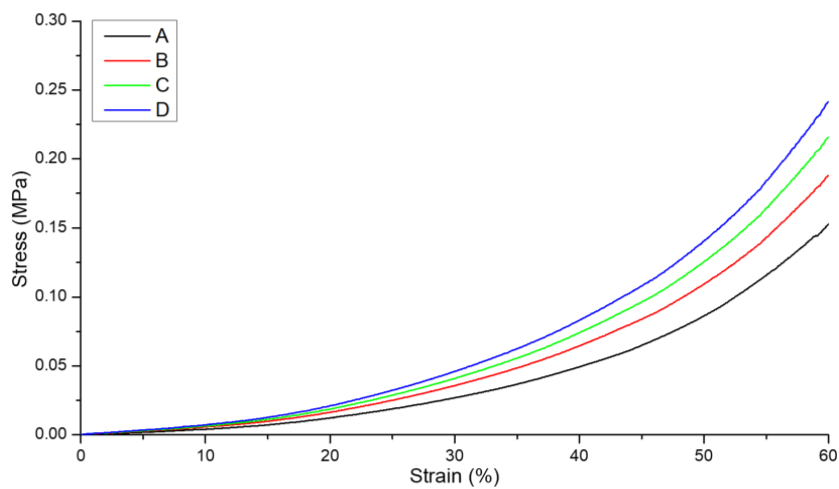

**Figure 11.** Initial linear elasticity portion of the compression stress-strain curve for the hydrogel and composite constructions. (A) 100 % PGSLPA hydrogel with no reinforcement. B) Bilayer nonwoven reinforced format. C) Trilayer nonwoven reinforced format D) Segmented nonwoven reinforced format.

**Table 3.** Compression strain and elastic modulus data for non-reinforced and nonwoven reinforced PGSLPA hydrogel samples.

| Structural Format | Polymer Composition | Concentration (wt/vol %) | Strain | E (kPa) |
|---|---|---|---|---|
| A<br>Fibre-free hydrogel | PGSLPA | 25 | 0.1 | 40 ± 28 |
| | | | 0.2 | 61 ± 13 |
| | | | 0.3 | 89 ± 21 |
| | PEGDA ($M_w$ 8000) | 10 | 0.4 | 123 ± 31 |
| | | | 0.5 | 172 ± 19 |
| | | | 0.6 | 254 ± 19 |
| B<br>Bilayer (hydrogel-nonwoven-hydrogel) | PGSLPA | 25 | 0.1 | 57 ± 41 |
| | | | 0.2 | 81 ± 21 |
| | PEGDA ($M_w$ 8000) | 10 | 0.3 | 119 ± 26 |
| | | | 0.4 | 161 ± 42 |
| | PGSLPA/PCL Electrospun nonwoven (2 g/m$^2$) | - | 0.5 | 218 ± 34 |
| | | | 0.6 | 313 ± 23 |
| C<br>Trilayer (hydrogel-nonwoven-hydrogel-nonwoven) | PGSLPA | 25 | 0.1 | 64 ± 44 |
| | | | 0.2 | 93 ± 20 |
| | PEGDA ($M_w$ 8000) | 10 | 0.3 | 136 ± 28 |
| | | | 0.4 | 184 ± 47 |
| | PGSLPA/PCL Electrospun nonwoven web (2 g/m$^2$) | - | 0.5 | 250 ± 37 |
| | | | 0.6 | 359 ± 12 |
| D<br>Fragmented (nonwoven reinforced hydrogel) | PGSLPA | 25 | 0.1 | 73 ± 53 |
| | | | 0.2 | 104 ± 28 |
| | PEGDA ($M_w$ 8000) | 10 | 0.3 | 153 ± 36 |
| | | | 0.4 | 207 ± 57 |
| | PGSLPA/PCL Electrospun nonwoven web (2 g/m$^2$) | 5 | 0.5 | 281 ± 49 |
| | | | 0.6 | 402 ± 38 |

The elastic modulus of the segmented nonwoven reinforced sample (format D) was significantly higher ($P<0.05$) than the unreinforced (format A) and bilayer (format B) constructions, at strain rates of 0.1 – 0.6, with the largest difference being between structural format types A and D.

Figure 12 illustrates the compression fatigue stress-strain curves (hysteresis) responses of hydrogel non-reinforced and reinforced structural formats A, B, C and D respectively. The hysteresis for the non-reinforced and reinforced PGSLPA hydrogels reveal clear elastomeric behaviour, with the deformation energy being absorbed, and samples returning to their original dimensions, as evidenced by the elastic recovery to 0% strain upon unloading. Although similar hysteresis response was observed for all samples, including the hysteresis response observed at 50 % maximum stress, at cycle 15, some differences were discernible. Cycle 1 showed an energy dissipation of 0.047, 0.041, 0.043 and 0.061, cycle 5 showed an energy dissipation of 0.029, 0.029, 0.028 and 0.040, cycle 15 showed an energy dissipation of 0.023,

0.021, 0.023 and 0.031 for format types A, B, C and D respectively.

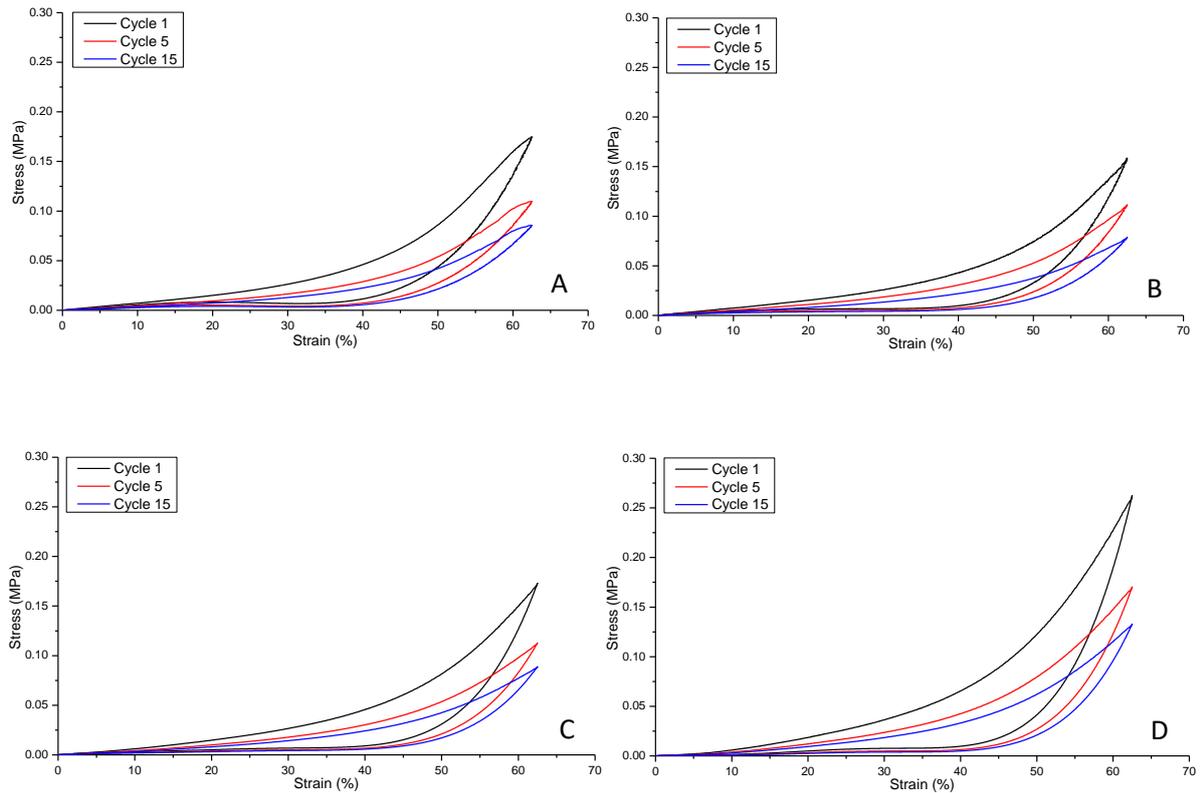

**Figure 12.** Compression fatigue stress-strain curves for: (A) unreinforced composite, A; (B) bilayer reinforced composite, B; (C) trilayer composite, C; (D) segmented electrospun nonwoven reinforced composite, D.

Comparison of the maximum stress at cycle 1 and cycle 5 of the samples suggests that the bilayer structure (Figure 12 B) produced the smallest decrease in maximum stress (30 %), compared to 37 %, and 35 % for formats A, and C and D respectively. However, cycle 15 revealed no significant differences ($P>0.05$) in terms of maximum stress, with decreases of 51 %, 51 %, 49% and 50 % for formats A, B, C and D respectively.

This behaviour is explained by the fibre orientation within the composite structure. Aligned nanofibres can be expected to impart anisotropic behaviour, such that the bulk structure exhibits preferential directional strength, compared to randomly oriented nanofibres that distribute the applied strain in an isotropic manner. The fibre diameter also plays an important role in determining the total surface area available for interfacial bonding with the surrounding hydrogel, affecting the overall mechanical strength of the composite. Physical contact between the fibre and hydrogel phases and interfacial bonding are essential to ensure efficient load transfer and prevent delamination or fibre pullout. Structural formats B and D (Figure 2) possessed similar fibre volume fractions, whereas for format C the volume fraction was slightly higher, which can lead to improved compressive strength compared to format B. However, the randomly segmented arrangement of nanofibres in structural format D is likely to explain the highest value for compressive recovery. Further improvements to the composite structure by combining the layering of nonwoven mats with an additional randomly orientated support phase, to increase the fibre volume fraction to >3% could improve mechanical stress responses still further.

## 4. Conclusions

Developing new elastomeric hydrogel materials capable of load bearing is important to enable tissue repair and regeneration in various clinical procedures, including for chronic wound care. To extend the potential clinical value of elastomeric PGS materials, it has been shown that a photocuring variant, poly(glycerol-*co*-sebacic acid-*co*-lactic acid-*co*-polyethylene glycol) acrylate (PGSLPA) can be synthesised and formed into elastomeric hydrogels, electrospun fibres and nonwovens as well as fibre-reinforced hydrogels, without the need for a high temperature curing step. The ability to photocure PGSLPA, improves the compatibility of PGS-based materials for use in minimally invasive clinical procedures and as part of injection treatments, coupled with photocuring *in situ*, obviating the need for the high temperature curing step, which is normally required. Photocuring of PGSLPA hydrogels can modulate the degradation rate, as well as bulk tensile and compression stress-strain behaviour, and elastomeric behaviour was observed in compression stress-strain evaluations. Reinforcement of PGSLPA hydrogels with a 50:50 PGSLPA/PCL nonwoven produced higher compressive stress and elastic moduli than unreinforced PGSLPA hydrogels. Sharp increases in stress values may be anticipated by increasing the fibre packing fraction beyond the 3% limit explored in this study, to engineer more resilient hydrogels for clinical procedures.

**Author Contributions:** Conceptualisation, S.R. and G.T.; Methodology, M.P., S.R. and G.T.; Software, M.P.; Validation, S.R., G.T., and C.P.; Formal Analysis, M.P.; Investigation, M.P.;

Resources, M.P. and C.P.; Data Curation, M.P.; Writing Original Draft Preparation, M.P.; Writing Review & Editing, S.R., G.T. and C.P.; Visualisation, M.P.; Supervision, S.R., G.T., and C.P.; Project Administration, S.R. and G.T.; Funding Acquisition, S.R. and G.T.

**Funding:** This research received external funding from The Clothworkers' Foundation, UK.

**Data Availability Statement:** Data are contained in the article.

**Acknowledgments:** The financial support of The Clothworkers' Foundation, which enabled a research scholarship to be provided for MP is gratefully acknowledged. The authors also wish to thank University of Leeds colleagues, Dr Zohreh Gharaei for help with electrospinning guidance, Dr Matthew Clark for his technical expertise, Dr Peter Broadbent for his help with experimental procedures, Dr Charles Brooker for help with compression testing guidance, and Dr Algy Kazlauciunas for generously providing the use of his Jeol JSM-6610LV scanning electron microcope. Mr. David Russell, Consultant Vascular Surgeon at Leeds Teaching Hospitals NHS Trust, UK is also gratefully acknowledged for his insights on the clinical challenges.

**Conflicts of Interest:** The authors declare no conflict of interest.